\documentclass[conference]{IEEEtran}
\IEEEoverridecommandlockouts
\usepackage{bibnames}
\usepackage{amssymb}
\usepackage{verbatim}
\usepackage{amsmath}
\usepackage{amsthm}
\usepackage[pdftex]{graphicx}
\usepackage[font=scriptsize]{caption}
\usepackage{subcaption}
 \usepackage[linesnumbered, ruled,vlined]{algorithm2e}
\usepackage{listings}
\usepackage[flushleft]{threeparttable}
\usepackage{multirow}
\usepackage{listings}
\usepackage{cite}
\usepackage{hyperref}
\usepackage{mdframed}
\usepackage{tabularx}
\usepackage{booktabs}
\usepackage{flushend}
\usepackage{mathrsfs}
\usepackage[T1]{fontenc}
\usepackage[switch]{lineno}
\usepackage{enumitem}

\usepackage{balance}

\definecolor{light-gray}{gray}{0.92}  
  {\begin{mdframed}[backgroundcolor=light-gray]\begin{mdtheorem}{name}{label}}%
  {\end{mdtheorem}\end{mdframed}}

\usepackage[table]{xcolor}
\definecolor{ao}{rgb}{0.0, 0.5, 0.0}
\newtheorem{definition}{Definition}

\newcommand{\etal}{\textit{et al.}\space}

\newcommand{\tool}{\textsc{VFFinder}\xspace}

\newcommand{\aAST}{\textsc{$\alpha$\textit{AST}}\xspace}

\newcommand{\aASTs}{\textsc{$\alpha$\textit{AST}s}\xspace}

\newcommand{\CEL}{CostEffort$@L$\xspace}

\newcommand{\jit}{just-in-time\space}
\def\BibTeX{{\rm B\kern-.05em{\sc i\kern-.025em b}\kern-.08em
    T\kern-.1667em\lower.7ex\hbox{E}\kern-.125emX}}
\begin{document}

\title{
\tool: A Graph-based Approach for\\Automated Silent Vulnerability-Fix Identification
}

\author{\IEEEauthorblockN{Son Nguyen, Thanh Trong Vu, and Hieu Dinh Vo$^*$\thanks{*Corresponding author.}}
\IEEEauthorblockA{\textit{Faculty of Information Technology} \\
\textit{University of Engineering and Technology, Vietnam National University, Hanoi, Vietnam} \\ \{sonnguyen,19020626,hieuvd\}@vnu.edu.vn} 
}

\maketitle

\begin{abstract}
The increasing reliance of software projects on third-party libraries has raised concerns about the security of these libraries due to hidden vulnerabilities. Managing these vulnerabilities is challenging due to the time gap between fixes and public disclosures. Moreover, a significant portion of open-source projects silently fix vulnerabilities without disclosure, impacting vulnerability management. Existing tools like OWASP heavily rely on public disclosures, hindering their effectiveness in detecting unknown vulnerabilities. To tackle this problem, automated identification of vulnerability-fixing commits has emerged. 
However, identifying silent vulnerability fixes remains challenging. This paper presents \tool, a novel graph-based approach for automated silent vulnerability fix identification. \tool captures structural changes using Abstract Syntax Trees (ASTs) and represents them in annotated ASTs. \tool distinguishes vulnerability-fixing commits from non-fixing ones using attention-based graph neural network models to extract structural features.
We conducted experiments to evaluate \tool on a dataset of 36K+ fixing and non-fixing commits in 507 real-world C/C++ projects. Our results show that \tool significantly improves the state-of-the-art methods by 39--83\% in Precision, 19--148\% in Recall, and 30\%--109\% in F1. Especially, \tool speeds up the silent fix identification process by up to 47\% with the same review effort of 5\% compared to the existing approaches.

\end{abstract}


\section{Introduction}
With the increasing reliance of software projects on third-party libraries, ensuring their security has become a paramount concern. Vulnerabilities hidden within these libraries can have far-reaching consequences, as exemplified by the infamous Log4Shell\footnote{\url{https://nvd.nist.gov/vuln/detail/CVE-2021-44228}} exploit. One critical challenge in addressing these vulnerabilities is the time gap between their fixes and public disclosures~\cite{iso-standard,cert-guideline}. 
For instance, Log4Shell's patch was pushed four days prior to its public revelation. 
Another example is that the Apache Struts Remote Code Execution vulnerability\footnote{\url{https://nvd.nist.gov/vuln/detail/CVE-2018-11776}}, which led to the Equifax breach in 2017, was disclosed to the public in August 2018, but was patched in June, 2018\footnote{\url{https://github.com/apache/struts/commit/6e87474}}.
Two months is plenty of time for the potential exploitation of vulnerable software.
If the library were monitored to identify vulnerability patches, the library's users would have been aware of the potential exploitation and prevented it by updating to the latest version of the component.

Despite the importance of the vulnerability fix identification task in open-source libraries, only a very small portion of maintainers file for a Common Vulnerability Enumeration (CVE) ID after releasing a fix, while 25\% of open-source projects silently fix vulnerabilities without disclosing them to any official repository~\cite{sspcatcher,snyk-report}. 
This situation raises concerns about the visibility and proactive management of vulnerabilities within the software ecosystem.
The open-source libraries' users rely on several tools and public vulnerability datasets like Open Web Application Security Project (OWASP) or National Vulnerability Database (NVD). However, CVE/NVD and public vulnerability databases miss many vulnerabilities~\cite{snyk-report}.

Recognizing the importance of identifying vulnerability-fixing commits, several security companies such as Huawei, Veracode, Mend, and Snyk have been monitoring open-source libraries' commits and other software artifacts to provide their users with early warnings of unpublished vulnerabilities.
However, the process of identifying silent vulnerability fixes is very challenging in practice. For example, constructing a dataset of 1,282 vulnerability-fixing commits required approximately four years~\cite{dataset-example}. Thus, automated vulnerability-fixing commit identification could help researchers maintain and update vulnerability databases, including NVD.

To address this problem, several vulnerability fix identification techniques have been proposed. Following the good practice of coordinated vulnerability disclosure~\cite{iso-standard,cert-guideline}, the related resources of commits, such as commit messages or issue reports, should not mention any security-related information before the public disclosure of the vulnerability. Thus, silent fix identification methods must not leverage these resources to classify commits. The state-of-the-art techniques, such as VulFixMiner~\cite{VulFixMiner}, CoLeFunDa~\cite{CoLeFunDa}, and Midas~\cite{Midas}, represent changes in the lexical form of code and apply CodeBERT~\cite{codebert} to capture code changes semantics and determine if they are vulnerability-fixing commit or not. Meanwhile, the existing studies have shown that the semantics of code changes could be captured better in the tree form of code~\cite{fira}.

In this paper, we propose \tool, a novel graph-based approach for automated vulnerability fix identification. Our idea is to capture the semantic meaning of code changes better, we explicitly represent the changes in code structure. Particularly, for a commit $c$, the structure of the versions before and after $c$ are represented by the Abstract Syntax Trees (ASTs). These ASTs are mapped to build an annotated AST (\aAST), a fine-grained graph representing the changes in the code structure caused by $c$. In \aASTs, all AST nodes and edges are annotated \textit{added}, \textit{deleted}, and \textit{unchanged} to explicitly express the changes in the code structure.
To learn the meanings of code changes expressed in \aASTs, we develop a graph attention network (GAT) model~\cite{gat} to extract semantic features. Then, these features are used to distinguish from vulnerability-fixing commits to non-fixing ones.

We conducted several experiments to evaluate \tool's performance on a dataset of 36K+ fixing and non-fixing commits in 507 real-world C/C++ projects. Our results show that \tool significantly improves the state-of-the-art techniques~\cite{Midas,VulFixMiner,jit-fine,jitline} by 39--83\% in Precision, up to 148\% in Recall, and 109\% in F1. Especially, \tool speeds up the silent fix identification process up to 47\%  with the same review effort of 5\% compared to the existing approaches.

In brief, this paper makes the following major contributions:

\begin{enumerate}
    \item {\tool}: A novel graph-based approach for identifying silent vulnerability-fixes.
    \item An extensive experimental evaluation showing the performance of {\tool} over the state-of-the-art methods for vulnerability-fix identification.
\end{enumerate}
%

\section{Code Change Representation}

In this work, we represent the syntactic aspect 
via the \textit{structure} relation using Abstract Syntax Tree (ASTs).

\begin{definition}[Annotated AST - \aAST]
For a commit changing code from one version to another, the \textit{annotated abstract syntax tree (annotated AST)} is an annotated graph constructed from the ASTs of these two versions. Formally, for $AST_o = \langle N_o, E_o \rangle$ and $AST_{n} = \langle N_n, E_n\rangle$ which are the ASTs of the old version and the new version, respectively, the \aAST $\mathcal{T} = \langle \mathcal{N}, \mathcal{E}, \alpha \rangle$ is defined as followings:
\begin{itemize}
    \item $\mathcal{N}$ is the set of the AST nodes in the old and new versions, $\mathcal{N} = N_o \cup N_n$.

    \item $\mathcal{E}$ is the set of the edges representing the structural relations between AST nodes in $AST_o$ and $AST_n$, $\mathcal{E} = E_o \cup E_n$.

    \item Annotations for nodes and edges in $\mathcal{T}$ are either \textit{unchanged}, \textit{added}, or \textit{deleted} by the change. Formally, $\alpha(g) \in \{$\textit{unchanged}, \textit{added}, \textit{deleted}$\}$, where $g$ is a node in $\mathcal{N}$ or an edge in $\mathcal{E}$:
        \begin{itemize}
            \item $\alpha(g) =$ \textit{added} if $g$ is a node and $g \in N_n \setminus N_o$, or $g$ is an edge and $g \in E_n \setminus E_o$
            \item $\alpha(g) =$ \textit{deleted} if $g$ is a node and $g \in N_o \setminus N_n$, or $g$ is an edge and $g \in E_o \setminus E_n$
            \item Otherwise, $\alpha(g) = $ \textit{unchanged} 
        \end{itemize}
\end{itemize}
\end{definition}

\begin{figure}
    \centering
    \includegraphics[width=0.8\columnwidth]{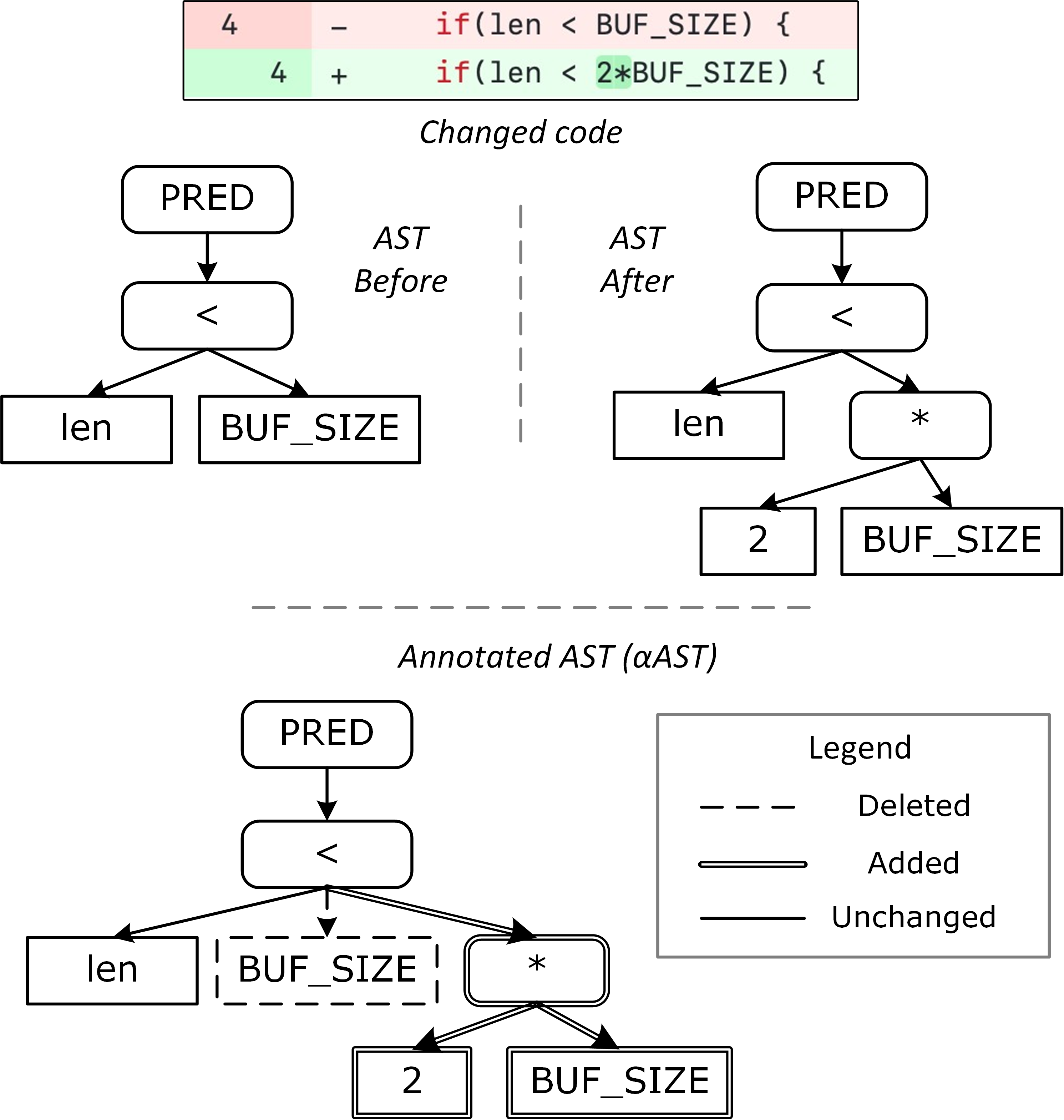}
    \caption{Annotated AST: An example}
    \label{fig:example}
\end{figure}

Fig.~\ref{fig:example} shows a piece of changed code, the ASTs before and after the change, and the annotated AST constructed from these ASTs. The \aAST expresses the change in the structure of the code at line 4. Particularly, the right-hand-side of the \texttt{less-than} expression (\texttt{BUF\_SIZE}) is replaced by the \texttt{multiply-expression} (i.e., \texttt{2*BUF\_SIZE}).



\section{Approach}
\begin{figure*}
    \centering
    \includegraphics[width=2\columnwidth]{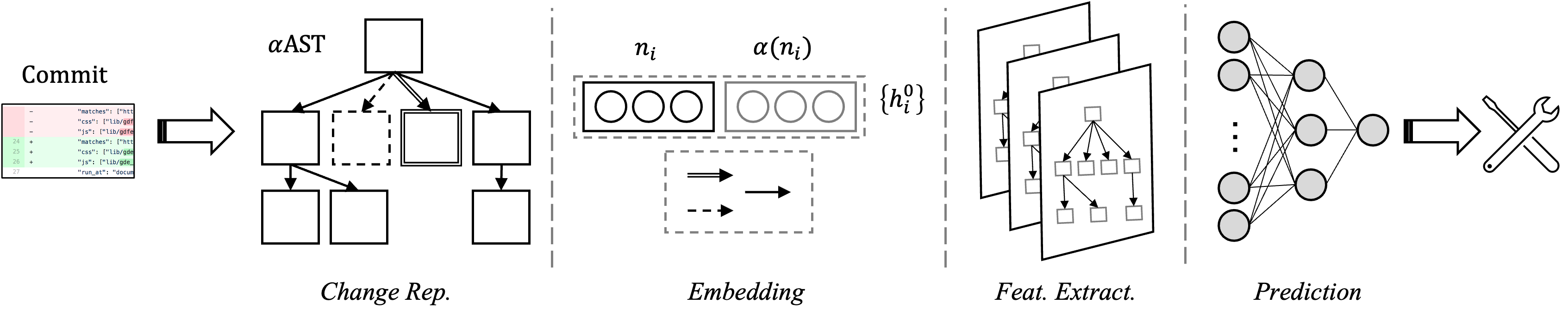}
    \caption{\tool: A Graph-based Approach for Vulnerability-fixing Commit Identification}
    \label{fig:approach}
\end{figure*}

Fig.~\ref{fig:approach} illustrates the overview of our approach, \tool, for vulnerability-fixing commit identification. First, the given commits and their repositories are used to construct their corresponding \aASTs (\textit{Change representation}). Each AST node in \aASTs is embedded in the corresponding vectors (\textit{Embedding}). After that, a Graph Neural Network (GNN) is applied to extract structural features from constructed \aASTs (\textit{Feature extraction}). Finally, the extracted structural features are used for learning and predicting vulnerability-fixing commits (\textit{Prediction}).

Particularly, in the \textit{Embedding} step, for each \aAST, $\mathcal{T} = \langle \mathcal{N}, \mathcal{E}, \alpha \rangle$, every node in $\mathcal{N}$ is embedded into $d$-dimensional hidden features $n_i$ produced by embedding the content of the nodes. 
To build the vectors for nodes' content, we use Word2vec~\cite{word2vec_1}, one of the most popular code embedding techniques for code~\cite{embedding_emse22}. The reason is that the number of AST nodes in \aASTs could be huge. Thus, for a practical embedding step for \aASTs, we apply Word2vec, which is known as an efficient embedding technique~\cite{embedding_emse22}.
Then, to form the node feature vectors, the node embedding vectors are annotated with the change operators (\textit{added}, \textit{deleted}, and \textit{unchanged}) by concatenating corresponding one-hot vector of the operators to the embedded vectors, $h^{0}_i = [n_i || \alpha(n_i)]$, where $||$ is the concatenation operation and $\alpha$ returns the one-hot vector corresponding the annotation of node $i$.

In the \textit{Feature Extraction} step, from each \aAST, $\mathcal{T} = \langle \mathcal{N}, \mathcal{E}, \alpha \rangle$, we develop a Graph Attention Network (GAT)~\cite{gat} model to extract the structural features $H$.
Particularly, the embedded vectors of the nodes from the \textit{Embedding} step are fed to a GAT model.
Each GAT layer computes the representations for the graph's nodes through message passing~\cite{gcn,gat}, where each node gathers features from its neighbors to represent the local graph structure. Stacking $L$ layers allows the network to build node representations from each node's $L$-hop neighborhood.
From the feature vector $h_{i}$ of node $i$ at the current layer, the feature vector $h'_{i}$ at the next layer is: 
$$
h'_{i} =\sigma 
        \left( 
            \sum_{j \in \mathcal{N}_i} \alpha_{ij} \text{\textbf{W}}h_{j}
        \right)
$$
where \textbf{W} is a learnable weight matrix for feature transformation, $\mathcal{N}_i$ is the set of neighbor indices of node $i$ including node $i$ itself via \textit{self-connection}, which is a single special relation from node $i$ to itself.
$\sigma$ is a non-linear activation function such as ReLU. Meanwhile, $\alpha_{ij}$ specifies the weighting factor (importance) of node $j$'s features to node $i$. $\alpha_{ij}$ could be explicitly defined based on the structural properties of the graph or learnable weight~\cite{gcn,fastgcn}. In this work, we implicitly define $\alpha_{ij}$ based on node features~\cite{gat} by employing the self-attention mechanism, where unnormalized coefficients $E_{ij}$ for pairs of nodes $i,j$ are computed based on their features:
$$
E_{ij} = \text{LeakyReLu}(\text{\textbf{a}}^T  \cdot [\text{\textbf{W}}h_i || \text{\textbf{W}}h_j] ), 
$$
where $||$ is the concatenation operation and \textbf{a} is a parametrizing weight vector implemented by a single-layer feed-forward neural network. $E_{ij}$ indicates the importance of node $j$'s features to node $i$. The coefficients are normalized across all choices of $j$ using the softmax function:
$$
\alpha_{ij} = \text{softmax}_j(E_{ij}) = 
\frac{\exp(E_{ij})}{\sum_{k \in \mathcal{N}_i} \exp(E_{ik}) }
$$
After $L$ GAT layers, a $d$-dimensional graph-level vector representation $H$ for the whole \aAST $\mathcal{T} = \langle \mathcal{N}, \mathcal{E}, \alpha \rangle$ is built by averaging over all node features in the final GAT layer, $H = \frac{1}{|\mathcal{N}|} \sum_{i \in [1, |\mathcal{N}|]} h^L_i$.
Finally, in the \textit{Prediction} step, the graph features are then passed to a Multilayer perceptron (MLP) to classify if \aAST $\mathcal{T}$ is a fixing commit or not.

\section{Evaluation Methodology}
To evaluate our vulnerability-fixing commit identification approach, we seek to answer the following research questions:

\noindent\textbf{RQ1: \textit{Accuracy and Comparison}.} How accurate is {\tool} in identifying vulnerability-fixing commits? And how is it compared to the state-of-the-art approaches~\cite{Midas,VulFixMiner}?

\noindent\textbf{RQ2: \textit{Sensitivity Analysis}.} How do various factors of the input, including training data size and changed code complexity, affect {\tool}'s performance?

\noindent\textbf{RQ3: \textit{Time Complexity}.} What is  {\tool}'s running time?

\subsection{Dataset}
In this work, we collect the vulnerability-fixing commits from various public vulnerability datasets~\cite{fixing_database1,bigvul,devign}.
In total, we collected the fixing commits for the vulnerabilities reported from 1990 to 2022 in real-world 507 C/C++ projects.
%
For a pragmatic evaluation, we collected the (remaining) non-fixing commits in the popular projects among these projects, such as FFmpeg, Qemu, Linux, and Tensorflow. 
The total numbers of collected fixing commits and non-fixing commits are about 11K and 25K, respectively.
Table~\ref{tab:dataset} shows the overview of our dataset. The details of our dataset can be found at: \textit{\url{https://github.com/thanhtlx/VFFinder}}.

\begin{table}[]
\centering
\caption{Dataset statistics}
\label{tab:dataset}
\begin{tabular}{@{}lrrrr@{}}
\toprule
         & \#\textit{Projects}   & \#\textit{Fixes}     & \#\textit{Non-fixes}       & \#\textit{Changed LOC}  \\ \midrule
Training & 402          & 9,176       & 21,193            & 3,807,967         \\
Testing  & 105          & 2,123       & 3,878             & 992,938         \\ \midrule
Total    & 507          & 11,299      & 25,071            & 4,800,905         \\ \bottomrule
\end{tabular}
\end{table}

\subsection{Procedure}
For \textbf{RQ1. Accuracy and Comparison}, we compared \tool against the state-of-the-art vulnerability fix identification approaches: 

1) \textbf{MiDas}~\cite{Midas} constructs different neural networks for each level of code change granularity, corresponding to commit-level, file-level, hunk-level, and line-level, following their natural organization. It then utilizes an ensemble model that combines all base models to generate the final prediction.

2) \textbf{VulFixMiner}~\cite{VulFixMiner} and CoLeFunDa~\cite{CoLeFunDa} utilize CodeBERT to automatically represent code changes and extract features for identifying vulnerability-fixing commits. However, as the implementation of CoLeFunDa has not been available, we cannot compare \tool with CoLeFunDa. This is also the reason that Zhou~\etal was not able to compare MiDas with CoLeFunDa in their study~\cite{Midas}.  

Additionally, we applied the same procedure adapting the state-of-the-art just-in-time defect detection techniques for vulnerability fix identification as in the work of Zhou~\etal~\cite{Midas}. The additional baselines include:

%
%

3) \textbf{JITLine}~\cite{jitline}: A simple but effective method utilizing changed code and expert features to detect buggy commits. 
%
%

%
%

4) \textbf{JITFine}~\cite{jit-fine}: A DL-based approach extracting features of commits from changed code and commit message using CodeBERT as well as expert features.

Note that we did not utilize commit messages when adapting JITLine and JITFine for silent vulnerability fix identification in our experiments. For \tool, we set the number of GAT layers $L=2$ for a practical evaluation.

In this comparative study, we follow the same cross-project evaluation procedure to construct the training data and testing data from the dataset as in the prior work~\cite{Midas,VulFixMiner}.
Particularly, the whole set of projects is randomly split into 80\% (402 projects) for training and 20\% (105 projects) for testing. The details of the training set and testing set are shown in Table~\ref{tab:dataset}.


For \textbf{RQ2. Sensitivity Analysis}, we studied the impacts of the following factors on \tool's performance: training size and change size in the number of changed lines of code (LOCs). To systematically vary these factors, we gradually added more training data and varied the change size.

\subsection{Metrics}
To evaluate the vulnerability fix identification approaches, we measure the classification \textit{accuracy}, \textit{precision}, and \textit{recall}, as well as \textit{F1}, which is a harmonic mean of precision and recall. 
Particularly, the classification accuracy (\textit{accuracy} for short) is the fraction of the (fixing and non-fixing) commits which are correctly classified among all the tested commits.
For detecting fixing commits, \textit{precision} is the fraction of correctly detected fixing commits among the detected fixing commits, while \textit{recall} is the fraction of correctly detected fixing commits among the fixing commits. Formally $precision = \frac{TP}{TP+FP}$ and $recall = \frac{TP}{TP+FN}$, where $TP$ is the number of true positives, $FP$ and $FN$ are the numbers of false positives and false negatives, respectively. \textit{F1} is calculated as $\textit{F1} = \frac{2 \times precision \times recall}{precision + recall}$.
Additionally, we also applied a cost-aware performance metric, \CEL (\textit{CE@L}), which is used in~\cite{Midas,VulFixMiner}. \textit{CE@L} counts the number of detected vulnerability-fixing commits, starting from commit with high to low predicted probabilities until the number of lines of code changes reaches L\% of total lines of code (LOC). 
%

\section{Experimental Results}

\subsection{Performance Comparison (RQ1)}
Table~\ref{tab:comparision} shows the performance of \tool and the state-of-the-art vulnerability fix identification approaches. As seen, \tool significantly outperforms the state-of-the-art vulnerability fix identification approaches. 
Particularly, the \tool achieves a recall of 0.57, which is more than \textbf{19--148\%} the recall rates of VulFixMiner and MiDas, respectively. Additionally, \tool is still much more precise than the existing approaches with about \textbf{39--83\%} improvement in the precision rate. 
These show that \tool can not only find more vulnerability-fixing commits but also provide much more precise predictions.
Furthermore, the \textit{CE@5\%} of \tool is \textbf{0.50}, which is \textbf{19--47\%} better than the corresponding figures of MiDas and VulFixMiner. This means that given 5\% effort (in LOC), the number of the fixing commits found by using \tool is much larger compared to those found by using MiDas and VulFixMiner.


\textbf{Answer to RQ1}: \tool is more effective than the state-of-the-art approaches in identifying vulnerability fixes. This confirms our strategy explicitly representing the code structure changes and using graph-based models to extract features for vulnerability fix identification.

\begin{table}[]
\centering
\caption{Comparison Results}
\label{tab:comparision}
\begin{tabular}{@{}lrrrrrr@{}}
\toprule
 & \textit{Pre.} & \textit{Rec}. & \textit{F1} & \textit{Acc.} & \textit{AUC} & \textit{CE@5\%} \\ \midrule
JITLine                         & 0.62 & 0.23 & 0.33 & 0.68 & 0.65 & 0.23  \\
JITFine                         & 0.58 & 0.48 & 0.53 & 0.69 & 0.69 & 0.45  \\ 
VulFixMiner                     & 0.47 & 0.42 & 0.44 & 0.63 & 0.64 & 0.34  \\
MiDas                           & 0.53 & 0.25 & 0.34 & 0.66 & 0.58 & 0.42  \\ 

\midrule

\tool            & \textbf{0.86} & \textbf{0.57} & \textbf{0.69} & \textbf{0.82} & \textbf{0.79} &  \textbf{0.50} \\ 
\bottomrule
\end{tabular}
\end{table}

\subsection{Sensitivity Analysis (RQ2)}

\begin{figure}
    \centering
    \includegraphics[width=.8\columnwidth]{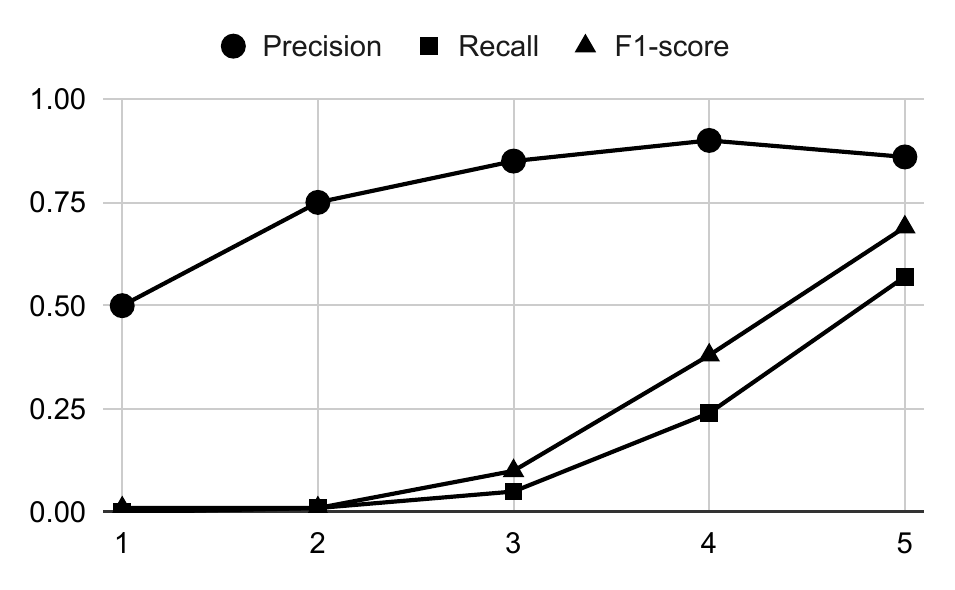}
    \caption{Impact of training data size on \tool's performance}
    \label{fig:impact_training_size}
\end{figure}

To measure the impact of training data size on \tool's performance. In this experiment, the training set is randomly separated into five folds.  We gradually increased the training data size by adding one fold at a time until all five folds were added for training.
As shown in Fig.~\ref{fig:impact_training_size}, \tool's performance is improved when expanding the training dataset. The precision increases by 72\% when the training data expands. Especially, the recall and F1-score grow much more significantly when increasing the training size from 1 fold to 5 folds. The reason is that with larger training datasets, \tool has observed more and then performs better. However, training with a larger dataset costs more time. The training time of \tool with five folds is about 3X more than that with a fold.

\begin{figure}
    \centering
    \includegraphics[width=1\columnwidth]{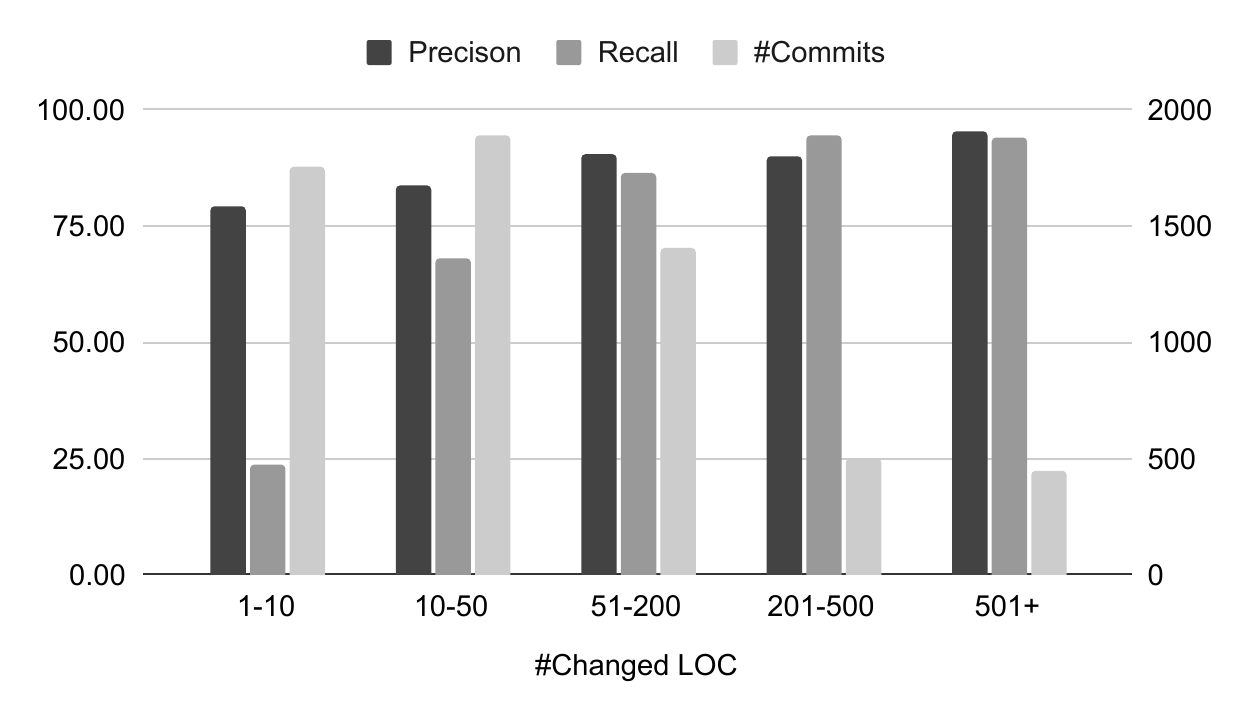}
    \caption{Impact of change size (left axis: \textit{Precision} and \textit{Recall}; right axis: No. of commits)}
    \label{fig:impact_change_size}
\end{figure}

Additionally, we investigate the sensitivity of \tool's performance on the input size in the number of changed (i.e., added and deleted) lines of code (LOC) (Fig.~\ref{fig:impact_change_size}). As seen, there are fewer commits with a larger number of changed LOC. 
The \textit{precision} of \tool is quite stable when handling commits in different change sizes. Meanwhile, the \textit{recall} significantly grows from 24\% to 93\% when increasing the change size. The reason could be that fixing commits tend to have less changed code~\cite{lapredict,jit-fine}. Thus, in the set of commit having a smaller number of changes, the vulnerability fix identification techniques could achieve a lower recall due to a larger number of fixing commits being identified.

\textbf{Answer for RQ2}: \tool performs better when being trained on a larger dataset. Additionally, \tool's precision is quite stable with different change sizes, while the recall is better with larger code changes.

\subsection{Time Complexity (RQ3)}
In this work, all our experiments were run on a server running Ubuntu 18.04 with an NVIDIA Tesla P100 GPU. 
In \tool, training the model took about 4--6 hours for 50 epochs. Additionally, \tool spent 1--2 seconds to classify whether a commit is a fixing commit or not. 
%


\subsection{Threats to Validity}
The main threats to the validity of our work consist of internal, construct, and external threats.

\textbf{Threats to internal validity} include the influence of the method used to construct AST. To reduce this threat, we use the widely-used code analyzer Joern~\cite{joern}. Another threat mainly lies in the correctness of the implementation of our approach. To reduce such a threat, we carefully reviewed our code and made it public~\cite{VFFinder} so that other researchers can double-check and reproduce our experiments.

\textbf{Threats to construct validity} relate to the suitability of our evaluation procedure. We used \textit{precision}, \textit{recall}, \textit{F1}, \textit{AUC}, \textit{accuracy}, and \CEL. They are the widely-used evaluation measures for vulnerability fix identification and \jit defect detection~\cite{jit-fine,jitline, VulFixMiner, Midas}. 
Besides, a threat may come from the adaptation of the baselines. To mitigate this threat, we directly obtain the original source code from their GitHub repositories or replicate exactly their description in the paper~\cite{VulFixMiner}. Also, we use the same hyperparameters as in the original papers~\cite{jit-fine,jitline,deepjit,Midas}. 

\textbf{Threats to external validity} mainly lie in the selection of graph neural network models used in our experiments. 
To mitigate this threat, we select the representative models which are well-known for NLP and SE tasks~\cite{gat, gcn}. 
%
Moreover, our experiments are conducted on only the code changes of C/C++ projects. Thus, the results could not be generalized for other languages. In our future work, we plan to conduct more experiments to validate our results in other languages. 
\section{Related Work}









\tool relates to the vulnerability fix identification work. VulFixMiner~\cite{VulFixMiner} and CoLeFunDa~\cite{CoLeFunDa} utilize CodeBERT to automatically represent code changes and extract features for identifying vulnerability-fixing commits.
Midas~\cite{Midas} constructs different neural networks for each level of code change granularity, corresponding to commit-level, file-level, hunk-level, and line-level, following their natural organization. It then utilizes an ensemble model that combines all base models to generate the final prediction.

\tool also relates to the work on just-in-time vulnerability detection.
DeepJIT~\cite{deepjit} automatically extracts features from commit messages and changed code and uses them to identify defects. 
Pornprasit~\etal propose JITLine, a simple but effective just-in-time defect prediction approach. JITLine utilizes the expert features and token features using bag-of-words from commit messages and changed code to build a defect prediction model with a random forest classifier. 
LAPredict~\cite{lapredict} is a defect prediction model by leveraging the information of ``lines of code added'' expert feature with the traditional logistic regression classifier. 
Recently, Ni~\etal introduced JITFine~\cite{jit-fine}, combining the expert features and the semantic features which are extracted by CodeBERT~\cite{codebert} from changed code and commit messages. 
Different from all prior studies in vulnerability fix identification and \jit bug detection, our work presents \tool which explicitly represents code changes in code structure and applies a graph-based model to extract the features distinguishing fixing commits from non-fixing ones. 


Several studies have been proposed for specific SE tasks, including code suggestion/completion~\cite{icse20, autosc,arist}, program synthesis~\cite{gvero2015synthesizing}, pull request description generation~\cite{hu2018deep,liu2019automatic}, code summarization~\cite{iyer2016summarizing,mastropaolo2021studying,wan2018improving}, code clones~\cite{li2017cclearner}, fuzz testing\cite{godefroid2017learn}, code-text translation~\cite{ase22}, bug/vulnerability detection~\cite{oppsla19,codejit, vultype}, and program repair~\cite{jiang2021cure,ding2020patching}.
\section{Conclusion}

%
In this paper, we have introduced \tool, an novel graph-based approach designed for the automated identification of vulnerability-fixing commits. By leveraging ASTs to capture structural changes and representing them as annotated ASTs, \tool enables the extraction of essential structural features. These features are then utilized by graph-based neural network models to differentiate vulnerability-fixing commits from non-fixing ones.
Our experimental results show that \tool improves the state-of-the-art methods by 30--60\% in Precision, 2.0X--4.0X in Recall, and 62\%--160\% in F1. Especially, \tool speeds up the silent fix identification process up to 2.6X with the same review effort of 5\%.
These findings highlight the superiority of \tool in accurately identifying vulnerability fixes and its ability to expedite the review process. The performance of \tool contributes to enhancing software security by empowering developers and security auditors with a reliable and efficient tool for identifying and addressing vulnerabilities in a timely manner.
\newpage

\bibliographystyle{IEEEtran}

\bibliography{main}

\begin{thebibliography}{10}
\providecommand{\url}[1]{#1}
\csname url@samestyle\endcsname
\providecommand{\newblock}{\relax}
\providecommand{\bibinfo}[2]{#2}
\providecommand{\BIBentrySTDinterwordspacing}{\spaceskip=0pt\relax}
\providecommand{\BIBentryALTinterwordstretchfactor}{4}
\providecommand{\BIBentryALTinterwordspacing}{\spaceskip=\fontdimen2\font plus
\BIBentryALTinterwordstretchfactor\fontdimen3\font minus
  \fontdimen4\font\relax}
\providecommand{\BIBforeignlanguage}[2]{{%
\expandafter\ifx\csname l@#1\endcsname\relax
\typeout{** WARNING: IEEEtran.bst: No hyphenation pattern has been}%
\typeout{** loaded for the language `#1'. Using the pattern for}%
\typeout{** the default language instead.}%
\else
\language=\csname l@#1\endcsname
\fi
#2}}
\providecommand{\BIBdecl}{\relax}
\BIBdecl

\bibitem{iso-standard}
\BIBentryALTinterwordspacing
{14:00-17:00}, ``\BIBforeignlanguage{en}{{ISO}/{IEC} 29147:2018}.'' [Online].
  Available: \url{https://www.iso.org/standard/72311.html}
\BIBentrySTDinterwordspacing

\bibitem{cert-guideline}
A.~D. Householder, G.~Wassermann, A.~Manion, and C.~King, ``The cert guide to
  coordinated vulnerability disclosure,'' \emph{Software Engineering Institute,
  Pittsburgh, PA}, 2017.

\bibitem{sspcatcher}
A.~D. Sawadogo, T.~F. Bissyand{\'e}, N.~Moha, K.~Allix, J.~Klein, L.~Li, and
  Y.~Le~Traon, ``Sspcatcher: Learning to catch security patches,''
  \emph{Empirical Software Engineering}, vol.~27, no.~6, p. 151, 2022.

\bibitem{snyk-report}
L.~Tal, ``The state of open source security report,'' Snyk, Tech. Rep., 2019.

\bibitem{dataset-example}
S.~E. Ponta, H.~Plate, A.~Sabetta, M.~Bezzi, and C.~Dangremont, ``A
  manually-curated dataset of fixes to vulnerabilities of open-source
  software,'' in \emph{2019 IEEE/ACM 16th International Conference on Mining
  Software Repositories (MSR)}.\hskip 1em plus 0.5em minus 0.4em\relax IEEE,
  2019, pp. 383--387.

\bibitem{VulFixMiner}
J.~Zhou, M.~Pacheco, Z.~Wan, X.~Xia, D.~Lo, Y.~Wang, and A.~E. Hassan,
  ``Finding a needle in a haystack: Automated mining of silent vulnerability
  fixes,'' in \emph{2021 36th IEEE/ACM International Conference on Automated
  Software Engineering (ASE)}.\hskip 1em plus 0.5em minus 0.4em\relax IEEE,
  2021, pp. 705--716.

\bibitem{CoLeFunDa}
J.~Zhou, M.~Pacheco, J.~Chen, X.~Hu, X.~Xia, D.~Lo, and A.~E. Hassan,
  ``Colefunda: Explainable silent vulnerability fix identification,'' 2023.

\bibitem{Midas}
T.~G. Nguyen, T.~Le-Cong, H.~J. Kang, R.~Widyasari, C.~Yang, Z.~Zhao, B.~Xu,
  J.~Zhou, X.~Xia, A.~E. Hassan \emph{et~al.}, ``Multi-granularity detector for
  vulnerability fixes,'' \emph{IEEE Transactions on Software Engineering},
  2023.

\bibitem{codebert}
Z.~Feng, D.~Guo, D.~Tang, N.~Duan, X.~Feng, M.~Gong, L.~Shou, B.~Qin, T.~Liu,
  D.~Jiang, and M.~Zhou, ``{C}ode{BERT}: A pre-trained model for programming
  and natural languages,'' in \emph{Findings of the Association for
  Computational Linguistics: EMNLP 2020}.\hskip 1em plus 0.5em minus
  0.4em\relax Online: Association for Computational Linguistics, Nov. 2020, pp.
  1536--1547.

\bibitem{fira}
J.~Dong, Y.~Lou, Q.~Zhu, Z.~Sun, Z.~Li, W.~Zhang, and D.~Hao, ``Fira:
  fine-grained graph-based code change representation for automated commit
  message generation,'' in \emph{Proceedings of the 44th International
  Conference on Software Engineering}, 2022, pp. 970--981.

\bibitem{gat}
P.~V. G. C.~A. Casanova, A.~R.~P. Lio, and Y.~Bengio, ``Graph attention
  networks,'' \emph{ICLR. Petar Velickovic Guillem Cucurull Arantxa Casanova
  Adriana Romero Pietro Li{\`o} and Yoshua Bengio}, 2018.

\bibitem{jit-fine}
C.~Ni, W.~Wang, K.~Yang, X.~Xia, K.~Liu, and D.~Lo, ``The best of both worlds:
  integrating semantic features with expert features for defect prediction and
  localization,'' in \emph{Proceedings of the 30th ACM Joint European Software
  Engineering Conference and Symposium on the Foundations of Software
  Engineering}, 2022, pp. 672--683.

\bibitem{jitline}
C.~Pornprasit and C.~K. Tantithamthavorn, ``Jitline: A simpler, better, faster,
  finer-grained just-in-time defect prediction,'' in \emph{2021 IEEE/ACM 18th
  International Conference on Mining Software Repositories (MSR)}.\hskip 1em
  plus 0.5em minus 0.4em\relax IEEE, 2021, pp. 369--379.

\bibitem{word2vec_1}
T.~Mikolov, K.~Chen, G.~Corrado, and J.~Dean, ``Efficient estimation of word
  representations in vector space,'' in \emph{1st International Conference on
  Learning Representations, {ICLR} 2013, Scottsdale, Arizona, USA, May 2-4,
  2013}, Y.~Bengio and Y.~LeCun, Eds., 2013.

\bibitem{embedding_emse22}
Z.~Ding, H.~Li, W.~Shang, and T.-H.~P. Chen, ``Can pre-trained code embeddings
  improve model performance? revisiting the use of code embeddings in software
  engineering tasks,'' \emph{Empirical Software Engineering}, vol.~27, no.~3,
  pp. 1--38, 2022.

\bibitem{gcn}
T.~N. Kipf and M.~Welling, ``Semi-supervised classification with graph
  convolutional networks,'' in \emph{International Conference on Learning
  Representations}, 2016.

\bibitem{fastgcn}
J.~Chen, T.~Ma, and C.~Xiao, ``Fastgcn: fast learning with graph convolutional
  networks via importance sampling,'' \emph{arXiv preprint arXiv:1801.10247},
  2018.

\bibitem{fixing_database1}
G.~Bhandari, A.~Naseer, and L.~Moonen, ``Cvefixes: automated collection of
  vulnerabilities and their fixes from open-source software,'' in
  \emph{Proceedings of the 17th International Conference on Predictive Models
  and Data Analytics in Software Engineering}, 2021, pp. 30--39.

\bibitem{bigvul}
J.~Fan, Y.~Li, S.~Wang, and T.~N. Nguyen, ``A c/c++ code vulnerability dataset
  with code changes and cve summaries,'' in \emph{Proceedings of the 17th
  International Conference on Mining Software Repositories}, 2020, pp.
  508--512.

\bibitem{devign}
Y.~Zhou, S.~Liu, J.~Siow, X.~Du, and Y.~Liu, ``Devign: Effective vulnerability
  identification by learning comprehensive program semantics via graph neural
  networks,'' \emph{Advances in neural information processing systems},
  vol.~32, 2019.

\bibitem{lapredict}
Z.~Zeng, Y.~Zhang, H.~Zhang, and L.~Zhang, ``Deep just-in-time defect
  prediction: how far are we?'' in \emph{Proceedings of the 30th International
  Symposium on Software Testing and Analysis}, 2021, pp. 427--438.

\bibitem{joern}
F.~Yamaguchi, N.~Golde, D.~Arp, and K.~Rieck, ``Modeling and discovering
  vulnerabilities with code property graphs,'' in \emph{2014 IEEE Symposium on
  Security and Privacy}.\hskip 1em plus 0.5em minus 0.4em\relax IEEE, 2014, pp.
  590--604.

\bibitem{VFFinder}
\BIBentryALTinterwordspacing
 [Online]. Available: \url{https://github.com/UETISE/VFFinder}
\BIBentrySTDinterwordspacing

\bibitem{deepjit}
T.~Hoang, H.~K. Dam, Y.~Kamei, D.~Lo, and N.~Ubayashi, ``Deepjit: an end-to-end
  deep learning framework for just-in-time defect prediction,'' in \emph{2019
  IEEE/ACM 16th International Conference on Mining Software Repositories
  (MSR)}.\hskip 1em plus 0.5em minus 0.4em\relax IEEE, 2019, pp. 34--45.

\bibitem{icse20}
S.~Nguyen, H.~Phan, T.~Le, and T.~N. Nguyen, ``Suggesting natural method names
  to check name consistencies,'' in \emph{2020 42nd International Conference on
  Software Engineering}.\hskip 1em plus 0.5em minus 0.4em\relax IEEE, 2020, pp.
  1372--1384.

\bibitem{autosc}
S.~Nguyen, T.~Nguyen, Y.~Li, and S.~Wang, ``Combining program analysis and
  statistical language model for code statement completion,'' in \emph{2019
  34th IEEE/ACM International Conference on Automated Software Engineering
  (ASE)}.\hskip 1em plus 0.5em minus 0.4em\relax IEEE, 2019, pp. 710--721.

\bibitem{arist}
S.~Nguyen, C.~T. Manh, K.~T. Tran, T.~M. Nguyen, T.-T. Nguyen, K.-T. Ngo, and
  H.~D. Vo, ``Arist: An effective api argument recommendation approach,''
  \emph{Journal of Systems and Software}, p. 111786, 2023.

\bibitem{gvero2015synthesizing}
T.~Gvero and V.~Kuncak, ``Synthesizing java expressions from free-form
  queries,'' in \emph{Proceedings of the 2015 ACM SIGPLAN International
  Conference on Object-Oriented Programming, Systems, Languages, and
  Applications}, 2015, pp. 416--432.

\bibitem{hu2018deep}
X.~Hu, G.~Li, X.~Xia, D.~Lo, and Z.~Jin, ``Deep code comment generation,'' in
  \emph{2018 IEEE/ACM 26th International Conference on Program Comprehension
  (ICPC)}.\hskip 1em plus 0.5em minus 0.4em\relax IEEE, 2018, pp. 200--20\,010.

\bibitem{liu2019automatic}
Z.~Liu, X.~Xia, C.~Treude, D.~Lo, and S.~Li, ``Automatic generation of pull
  request descriptions,'' in \emph{34th IEEE/ACM International Conference on
  Automated Software Engineering}.\hskip 1em plus 0.5em minus 0.4em\relax IEEE,
  2019, pp. 176--188.

\bibitem{iyer2016summarizing}
S.~Iyer, I.~Konstas, A.~Cheung, and L.~Zettlemoyer, ``Summarizing source code
  using a neural attention model,'' in \emph{Proceedings of the 54th Annual
  Meeting of the Association for Computational Linguistics (Volume 1: Long
  Papers)}, 2016, pp. 2073--2083.

\bibitem{mastropaolo2021studying}
A.~Mastropaolo, S.~Scalabrino, N.~Cooper, D.~N. Palacio, D.~Poshyvanyk,
  R.~Oliveto, and G.~Bavota, ``Studying the usage of text-to-text transfer
  transformer to support code-related tasks,'' in \emph{2021 IEEE/ACM 43rd
  International Conference on Software Engineering (ICSE)}.\hskip 1em plus
  0.5em minus 0.4em\relax IEEE, 2021, pp. 336--347.

\bibitem{wan2018improving}
Y.~Wan, Z.~Zhao, M.~Yang, G.~Xu, H.~Ying, J.~Wu, and P.~S. Yu, ``Improving
  automatic source code summarization via deep reinforcement learning,'' in
  \emph{Proceedings of the 33rd ACM/IEEE International Conference on Automated
  Software Engineering}, 2018, pp. 397--407.

\bibitem{li2017cclearner}
L.~Li, H.~Feng, W.~Zhuang, N.~Meng, and B.~Ryder, ``Cclearner: A deep
  learning-based clone detection approach,'' in \emph{International Conference
  on Software Maintenance and Evolution}.\hskip 1em plus 0.5em minus
  0.4em\relax IEEE, 2017, pp. 249--260.

\bibitem{godefroid2017learn}
P.~Godefroid, H.~Peleg, and R.~Singh, ``Learn\&fuzz: Machine learning for input
  fuzzing,'' in \emph{2017 32nd IEEE/ACM International Conference on Automated
  Software Engineering (ASE)}.\hskip 1em plus 0.5em minus 0.4em\relax IEEE,
  2017, pp. 50--59.

\bibitem{ase22}
H.~A. Nguyen, H.~D. Phan, S.~S. Khairunnesa, S.~Nguyen, A.~Yadavally, S.~Wang,
  H.~Rajan, and T.~Nguyen, ``A hybrid approach for inference between behavioral
  exception api documentation and implementations, and its applications,'' in
  \emph{37th IEEE/ACM International Conference on Automated Software
  Engineering}, 2022, pp. 1--13.

\bibitem{oppsla19}
Y.~Li, S.~Wang, T.~N. Nguyen, and S.~Van~Nguyen, ``Improving bug detection via
  context-based code representation learning and attention-based neural
  networks,'' \emph{Proceedings of the ACM on Programming Languages}, vol.~3,
  no. OOPSLA, pp. 1--30, 2019.

\bibitem{codejit}
S.~Nguyen, T.-T. Nguyen, T.~T. Vu, T.-D. Do, K.-T. Ngo, and H.~D. Vo,
  ``Code-centric learning-based just-in-time vulnerability detection,''
  \emph{arXiv preprint arXiv:2304.08396}, 2023.

\bibitem{vultype}
H.~D. Vo and S.~Nguyen, ``Can an old fashioned feature extraction and a
  light-weight model improve vulnerability type identification performance?''
  \emph{Information and Software Technology}, vol. 164, p. 107304, 2023.

\bibitem{jiang2021cure}
N.~Jiang, T.~Lutellier, and L.~Tan, ``Cure: Code-aware neural machine
  translation for automatic program repair,'' in \emph{2021 IEEE/ACM 43rd
  International Conference on Software Engineering (ICSE)}.\hskip 1em plus
  0.5em minus 0.4em\relax IEEE, 2021, pp. 1161--1173.

\bibitem{ding2020patching}
Y.~Ding, B.~Ray, P.~Devanbu, and V.~J. Hellendoorn, ``Patching as translation:
  the data and the metaphor,'' in \emph{2020 35th IEEE/ACM International
  Conference on Automated Software Engineering}.\hskip 1em plus 0.5em minus
  0.4em\relax IEEE, 2020, pp. 275--286.

\end{thebibliography}
\balance
\end{document}